# Late stages of mineralisation and their signature on the bone mineral density distribution

Pascal R. Buenzli[a,b,*], Chloé Lerebours[b], Andreas Roschger[c], Paul Roschger[d], Richard Weinkamer[c]

[a]School of Mathematical Sciences, Queensland University of Technology, Brisbane, Australia
[b]School of Mathematical Sciences, Monash University, Clayton, Australia
[c]Max Planck Institute of Colloids and Interfaces, Department of Biomaterials, Potsdam, Germany
[d]Ludwig Boltzmann Institute of Osteology at the Hanusch Hospital of WGKK and AUVA Trauma Centre Meidling, 1st Medical Department, Hanusch Hospital, Vienna, Austria
[*]pascal.buenzli@qut.edu.au

## Abstract

**Purpose**: Experimental measurements of bone mineral density distributions (BMDDs) enable a determination of secondary mineralisation kinetics in bone, but the maximum degree of mineralisation and how this maximum is approached remain uncertain. We thus test computationally different hypotheses on late stages of bone mineralisation by simulating BMDDs in low turnover conditions.
**Materials and Methods**: An established computational model of the BMDD that accounts for mineralisation and remodelling processes was extended to limit mineralisation to various maximum calcium capacities of bone. Simulated BMDDs obtained by reducing turnover rate from the reference trabecular BMDD under different assumptions on late stage mineralisation kinetics were compared with experimental BMDDs of low-turnover bone.
**Results**: Simulations show that an abrupt stopping of mineralisation near a maximum calcium capacity induces a pile-up of minerals in the BMDD statistics that is not observed experimentally. With a smooth decrease of mineralisation rate, imposing low maximum calcium capacities helps to match peak location and width of simulated low turnover BMDDs with peak location and width of experimental BMDDs, but results in a distinctive asymmetric peak shape. No tuning of turnover rate and maximum calcium capacity was able to explain the differences found in experimental BMDDs between trabecular bone (high turnover) and femoral cortical bone (low turnover).
**Conclusions**: Secondary mineralisation in human bone does not stop abruptly, but continues slowly up to a calcium content greater than 30 wt% Ca. The similar mineral heterogeneity seen in trabecular and femoral cortical bones at different peak locations was unexplained by turnover differences tested.

## Introduction

The mineralisation of newly formed bone occurs rapidly during the first few days, and then continues for months at a much slower rate in what is termed secondary mineralisation. With the continual replacement of bone tissue portions by new bone packets during remodelling, bone tissues thereby possess a high degree of mineral heterogeneity. A measure of this heterogeneity is captured by bone mineral density distributions (BMDDs), which is a frequency distribution of calcium content in bone [1]. The BMDD depends both on the mineralisation kinetics and on the remodelling process, and it can be measured from quantitative back-scattered electron microscopy scans of bone samples [1].

The kinetics of long-term secondary mineralisation is very hard to measure directly, but it was deduced indirectly from measurements of the trabecular BMDD of adults [2]. Mineral accumulation was found to continue steadily with a nonzero mineralisation rate for more than 30 years (assuming a turnover time of 5 years for trabecular bone) (Fig. 1). However, the mineralisation rate deduced from trabecular BMDDs in Ref [2] may be unreliable at high calcium content (i.e., at very late times). Voxels of high mineral content may comprise cement lines, mineralised cartilage, and hypermineralised defects that do not accumulate minerals in the same way as bone structural units. Also qBEI scans possess few voxels with high calcium content, which results in low counting statistics, and thus higher uncertainty [3].

Secondary mineralisation is thought to stop once all the non-bound water molecules in bone matrix are replaced by minerals. Consequently, mineralisation should only proceed up until a maximum calcium capacity of bone $c_{\max}$. Based on a collagen fibril model with its gap and overlap regions, the maximum possible volume fraction of the mineral phase in lamellar bone was estimated to be 0.56 [4], which would correspond to a calcium content of about 30 Ca wt% assuming minerals consisting of pure hydroxyapatite. However, an accurate estimation of the value of $c_{\max}$ in different mineralised tissues has to consider the contributions of both intrafibrillar and extrafibrillar minerals [5].

The aim of this work is to test different hypotheses about the very late stages of bone mineralisation using computer modelling. We investigate both the influence of different values of the maximum calcium capacity $c_{\max}$, as well as how fast this value may be reached. Various physical and biological factors may influence these parameters, including the permeability of the collagen fibril structure through which minerals diffuse, and regulations by osteocytes [6, 7]. It remains an open question whether minerals in a new bone packet keep accumulating at an increasingly reduced rate over very long times commensurate with the human lifespan, or if mineralisation stops earlier and therefore more abruptly, when $c_{\max}$ is reached. We test this question by simulating mineralisation rates in which mineralisation stops either smoothly or abruptly as the maximum calcium capacity $c_{\max}$ is reached. We calculate the BMDDs that result from these mineralisation rates and compare them with experimentally measured BMDDs. Situations of low bone turnover are particularly appropriate for studying the influence of late stages of mineralisation since the age of bone matrix is higher in these situations. Reduced turnover gives bone more time to mineralise before being remodelled, so that more bone packets may approach the maximum capacity $c_{\max}$ [8]. An example of low bone turnover is the effect of the administration of bisphosphonates [9]. Cortical bone is also an example of low bone turnover with a reduction of turnover rate by roughly a factor of 5 compared to trabecular bone [10]. It has been shown that cortical bone has a higher average bone mineral density than trabecular bone [11].

Earlier computational work demonstrated that a reduction of turnover rate in trabecular bone results in the long term in a broadening of the BMDD peak [2,12]. However, measurements of the BMDD in femoral cortical bone show that peak width $Ca_{WIDTH}$ is comparable with peak width of the reference trabecular BMDD. A specific aim of the study is thus to test whether peak broadening induced by turnover rate reduction can be counteracted by low values of $c_{\max}$. Mineralising bone packets at late stages of mineralisation may 'pile-up' against $c_{\max}$ in the BMDD statistics. This could provide a possible explanation for why the BMDD peak has a similar width in both trabecular and cortical bone.

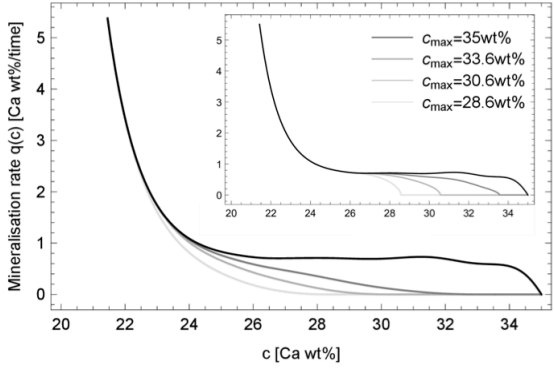 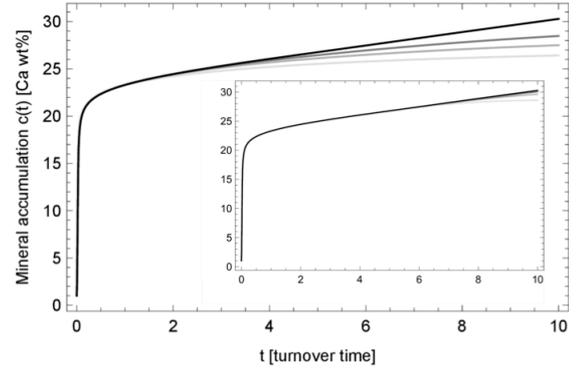

**Figure 1** – (a) Mineralisation rate $q(c)$ obtained from Eq. (1) (solid black lines) and mineralisation rates $q_{new}(c)$ obtained by multiplying $q(c)$ with smooth or abrupt step functions (grey lines); (b) Mineralisation kinetic laws c(t) obtained by solving $\frac{dc}{dt} = q(c)$ (solid black lines) or $\frac{dc}{dt} = q_{new}(c)$ (grey lines). The insets in (a) and (b) correspond to the situation of abrupt stopping of the mineralisation kinetics at $c_{max}$. The full-sized plots correspond to the situation of smooth stopping of the mineralisation kinetics at $c_{max}$. The maximum calcium capacities assumed in $q_{new}(c)$ are 33.6 wt% Ca, 30.6 wt% Ca, and 28.6 wt% Ca.

## Materials and Methods

The experimental data we consider are the reference trabecular BMDD obtained as the average of 52 BMDDs from healthy adults of both sexes and different skeletal sites [13(9)], as well as recent measurements of BMDDs in femoral cortical bone samples collected post mortem from four healthy women of age 48, 50, 55, and 56. The procedure to extract BMDDs as a function of calcium weight percent (wt% Ca) using quantitative backscattered electron imaging (qBEI) is detailed in Refs [1,14].

The BMDD is thought to be the outcome of (i) a continuous turnover of bone, and (ii) a mineralisation process that results in a monotonous increase of mineral content in a new bone packet starting from an initial unmineralised state. This view was translated in mathematical terms in Ref. [2], where the BMDD is represented as the solution of a reaction–advection equation. Advection along the mineral axis represents the increase in the degree of mineralisation of a bone packet due to mineralisation. The mineralisation rate $q(c)$ is defined such that a bone packet increases its calcium content $c$ by the amount $\Delta c = q(c)\Delta t$ during a small time increment $\Delta t$. The time course of mineral accumulation in a bone packet, $c(t)$, is such that $\frac{dc}{dt} = q(c)$. Reaction represents the removal of bone packets due to bone resorption occurring at rate $r$. Bone formation is represented as a boundary condition of the BMDD at $c = 0$, which corresponds to the mineral content of unmineralised matrix. In this study we consider experimental BMDDs of healthy adults only, and so we assume that these BMDDs are well described by a steady-state situation in which there is no temporal change in average mineral distribution and an approximate balance in the amount of bone that is resorbed and formed. In this situation, $r$ corresponds to the turnover rate (fraction of the bone volume BV under consideration that is renewed per unit time [15]). The quantity $1/r$ corresponds to the turnover time, i.e., the average time required to turn over BV [2].

The mathematical relationship between the steady state BMDD, the mineralisation rate $q(c)$, and the turnover rate $r$, can be inverted to determine the mineralisation rate $q(c)$ from the experimental reference trabecular BMDD as [2]:

$$q(c) = \frac{1}{\text{BMDD}(c)} \int_c^{c_{max}^{exp}} r\, \text{BMDD}(c')\, dc' \qquad (1)$$

where $c_{\max}^{\exp} = 35\text{wt\%}$ Ca is defined as the calcium concentration value represented by the last bin (at the high end) covered by the experimental BMDD histogram [1]. The mineralisation rate and mineralisation kinetic law obtained from Eq. (1) for (healthy) adult trabecular bone are shown in Figure 1. The mineralisation rate $q(c)$ reaches a plateau at a low, but nonzero value in the range 26wt%–34wt%. However, since measurements also include highly mineralised material elements that do not necessarily correspond to mineralising bone structural units and due to low electron counting statistics in the experiment, the derived mineralisation kinetics may not be reliable for these late stages of mineralisation.

To take into account this uncertainty about late stage mineralisation, we modify the mineralisation kinetics theoretically by hypothesising that mineralisation stops at lower values of $c_{\max}$ than $c_{\max}^{\exp}$. We tested three different values: $c_{\max} = 33.6$ wt% Ca, 30.6 wt% Ca, and 28.6 wt% Ca. We also allowed in the model for two different ways that $c_{\max}$ may be approached, either by a smooth decrease of the mineralisation rate towards zero, or by an abrupt stopping of mineral incorporation at $c_{\max}$. To implement these decreases of mineralisation rate $q(c)$ at $c_{\max}$ we multiply the mineralisation rate $q(c)$ in Eq. (1) with step functions $s(\frac{c-c_{\max}}{w})$ decreasing to zero in a transition region $c_{\max} - w < c \leq c_{\max}$, such that the modified mineralisation rates are $q_{\text{new}}(c) = q(c)s(\frac{c-c_{\max}}{w})$. For smooth decreases, the functions $s(x)$ are chosen to be fifth degree polynomials in the transition region, i.e., $s(x) = 1$ for $x \leq -1$, $s(x) = -x^3(10 + 15x + 6x^2)$ for $-1 < x < 0$, and $s(x) = 0$ for $x \geq 0$. The widths of the transitions are adjusted to prevent a plateau in mineralisation rate in the range 26wt%–34wt%. The modified mineralisation rates $q_{\text{new}}(c)$ are chosen to decrease smoothly to zero at $c_{\max} = 33.6$ wt% ($w = 11.11$ wt%), 30.6 wt% ($w = 8.33$ wt%), and 28.6 wt% ($w = 7.14$ wt%). For an abrupt stopping of mineralisation, the function $s(x)$ is chosen so that $q_{\text{new}}(c)$ decreases to zero as fast as $(c_{\max} - c)^{1/2}$ as $c$ approaches $c_{\max}$.

The original mineralisation rate $q(c)$ and the modified mineralisation rates $q_{\text{new}}(c)$ are shown in Fig. 1a. The corresponding mineralisation kinetics $c(t)$, obtained by solving $\frac{dc}{dt} = q_{\text{new}}(c)$, are shown in Fig. 1b. With these modified mineralisation rates, steady-state BMDDs are calculated as

$$\text{BMDD}(c) = \frac{r}{q_{\text{new}}(c)} \exp\left(-r \int_0^c \frac{1}{q_{\text{new}}(c')} dc'\right) \qquad (2)$$

(see Ref. [2]).

We simulated scenarios of low turnover by reducing the value of $r$ from the value 1, corresponding to normal turnover rate, to values as low as 0.06. The BMDDs obtained by this procedure were characterised by the following shape parameters commonly reported experimentally: $\text{Ca}_{\text{PEAK}}$, the calcium content for which the BMDD is maximum; $\text{Ca}_{\text{WIDTH}}$, the width of the BMDD peak at half height; and $\text{Ca}_{\text{HIGH}}$, which corresponds to the percentage of bone with a calcium content above 25.3 wt% Ca [1]. To measure the asymmetry of the BMDD peak, we additionally define a new parameter, $\text{Ca}_{\text{ASYM}}$, defined as the ratio of bone volume with calcium density higher than $\text{Ca}_{\text{PEAK}}$ and bone volume with calcium density lower than $\text{Ca}_{\text{PEAK}}$:

$$\text{Ca}_{\text{ASYM}} = \frac{\int_{\text{Ca}_{\text{PEAK}}}^{c_{\max}} \text{BMDD}(c) dc}{\int_0^{\text{Ca}_{\text{PEAK}}} \text{BMDD}(c) dc}. \qquad (3)$$

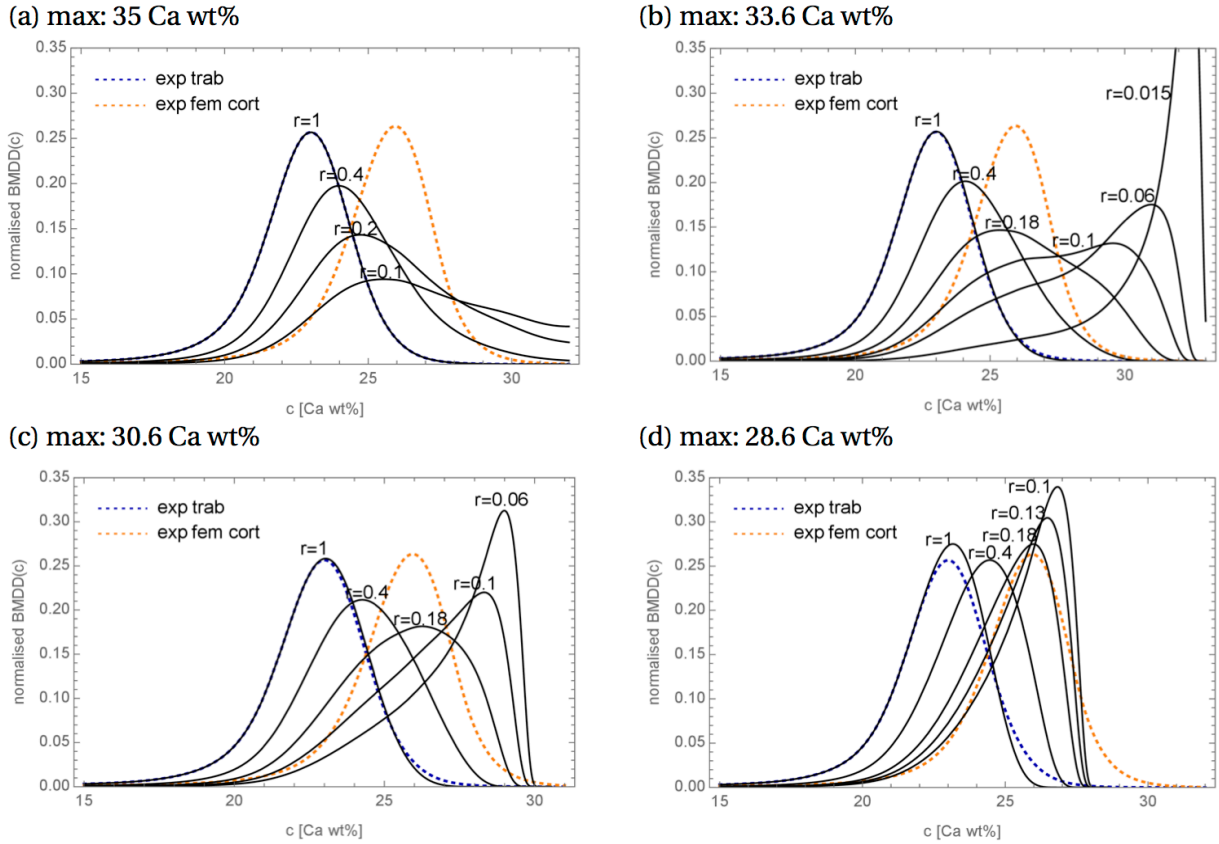

**Figure 2** - Bone mineral density distributions. Interrupted lines correspond to the reference trabecular BMDD (blue) and the femoral cortical BMDD (orange). Solid lines correspond to BMDDs calculated with mineralisation rates vanishing smoothly at $c_{max}$ as shown in Fig. 1a, and turnover rates reductions $r$, as indicated by the labels. (a) No correction in mineralisation rate, the maximum calcium capacity corresponds to the maximal experimental value measured, i.e., 35 wt% Ca; (b)–(d): mineralisation kinetics limited by a maximum calcium capacity of (b) 33.6 wt% Ca; (c) 30.6 wt% Ca; and (d) 28.6 wt% Ca.

## Results

Figures 2 and 3 shows the predicted steady-state BMDDs for different values of the maximum calcium capacity of bone $c_{max}$ and for either a smooth or abrupt stopping of the mineralisation process. For a smooth decrease (Fig. 2) the modification of the mineralisation rate due to a lowering of $c_{max}$ has the effect of lowering the high-calcium tail of the BMDD compared to the experimental reference trabecular BMDD, even for normal turnover rate ($r = 1$) (blue dotted line in Fig. 2). This effect is small except in the case of $c_{max} = 28.6 \text{wt\%}$ (Fig 2d), where this high-calcium tail lowering results in Ca$_{HIGH}$ decreasing from 5.55±3.32% (the reference value for trabecular bone [1]) to 1.41%. For an abrupt decrease of mineralisation rate (Fig. 3), the mineral density distribution piles up strongly near the value of maximum calcium capacity.

With decreasing turnover rate, the simulated steady-state BMDDs generally shift towards higher calcium densities, and the BMDD peaks broaden, as also observed in Ref. [2]. Figure 2 shows additionally that as the BMDD peak approaches the maximum calcium capacity in cases of low turnover, it changes shape dramatically due to a pile-up effect. This is particularly obvious in Figs 2b and 2c, where the BMDD peak transitions quickly (at around $r = 0.2-0.1$) from the left hand side of the distribution to the right hand side of the distribution. This transition occurs also in the uncorrected case (Fig. 2a) but at lower values of $r$ (not shown). The sharpness of this transition is lessened by the choice of a low maximum calcium capacity (Fig. 2d).

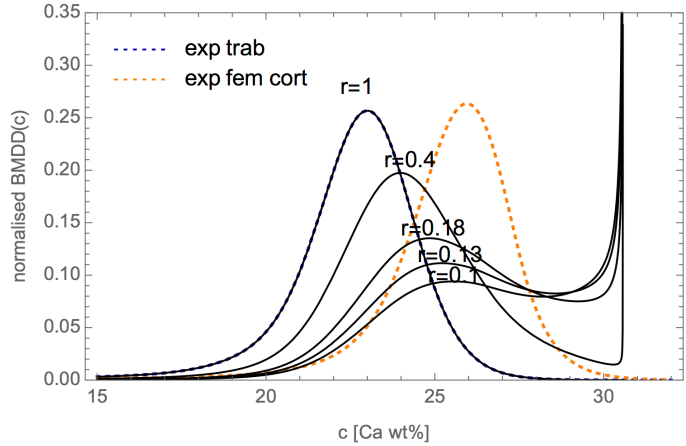

**Figure 3** – Bone mineral density distributions calculated with a mineralisation rate that vanishes abruptly at $c_{max}$ = 30.6 wt% Ca.

To highlight these transitions in peak shape we plotted in Fig. 4 the shape parameters $Ca_{PEAK}$, $Ca_{WIDTH}$, and $Ca_{ASYM}$ of the simulated BMDDs as functions of turnover rate $r$ alongside values measured on the experimental BMDDs (Table 1). The reduction of turnover rate results in an increase in $Ca_{PEAK}$ for all values of $c_{max}$ (Fig. 4a). However, the other two shape parameters $Ca_{WIDTH}$ and $Ca_{ASYM}$ follow more complex patterns as $r$ is reduced. They can take values above and below the experimental reference values. With decreasing turnover rate, both parameters first increase, before they decrease at low values of $r$. The sharpness of this transition is most pronounced for $c_{max}$ = 33.6wt% and is clearly smoothened by assuming lower maximum calcium capacities.

With this parameter analysis at hand, we can now tackle the problem of whether the cortical BMDD can be obtained from the trabecular reference BMDD by applying specific values of $c_{max}$ and turnover rate reduction $r$. The trabecular and cortical BMDDs determined by qBEI mostly differ by their value of $Ca_{PEAK}$. The parameters $Ca_{WIDTH}$ and $Ca_{ASYM}$ are very similar (Table 1). Lowering the maximum calcium capacity is effective to prevent peak broadening due to the pile-up effect. With $c_{max}$ = 28.6wt%, $Ca_{WIDTH}$ changes little compared to higher values of $c_{max}$ as $r$ is reduced (Fig. 4b). To match both $Ca_{PEAK}$ and $Ca_{WIDTH}$ of the cortical femoral BMDD, a reduction in turnover rate $r \approx 0.18$ at maximum calcium capacity $c_{max}$ = 28.6wt% seems effective. The BMDD comes close to the femoral cortical BMDD, particularly to the left of the peak (Fig. 2d). However, this low maximum calcium capacity significantly skews the peak shape, so that the BMDD deviates from the femoral cortical BMDD substantially to the right of the peak. Figure 4c shows that $Ca_{ASYM}$ at $r$ = 0.18, $c_{max}$ = 28.6wt% lies as far as 51% of the experimental cortical value.

**Table 1**: Shape parameters of the BMDD peaks (mean ± standard deviation). Values for the reference trabecular BMDD are from Ref. [1].

|  | $Ca_{PEAK}$ [wt%] | $Ca_{WIDTH}$ [wt%] | $Ca_{ASYM}$ [-] |
|---|---|---|---|
| reference trabecular BMDD | 22.20 ± 0.45 | 3.35 ± 0.22 | 0.80 ± 0.14 |
| femoral cortical BMDD | 25.95 ± 0.22 | 3.20 ± 0.10 | 0.74 ± 0.04 |

## Discussion

Our computational study allows us to analyse how dynamic parameters characterising remodelling and mineralisation influence parameters characterising BMDD peak shape. Motivated by the uncertainty of the kinetics of late stages of mineralisation, we investigated the influence on BMDDs of the maximum amount of calcium that can be incorporated in a small volume of bone matrix.

The results of our study lead us to the following four conclusions:
1) The assumption that bone mineralisation stops at a rather low value of calcium content in trabecular bone results in consequences that are in disagreement with experimental observations. The choice of $c_{max} = 28.6$ wt% causes a roughly four-times lower value of $Ca_{HIGH}$ than the reported reference value already in the case without turnover rate reduction (Fig. 2c, $r = 1$). Therefore, we conclude that mineralisation continues in trabecular bone – at least slowly – beyond this value.
2) An abrupt stopping of the mineralisation process at the maximum calcium capacity enables $c_{max}$ to be reached in finite time, but it leads to a sharp spike in the BMDD close the maximum calcium capacity $c_{max}$. Being aware of the absence of biological variance and experimental noise in the computer simulation, which makes the simulated peak particularly spiked (Fig. 3), we are nonetheless confident in stating that an abrupt end of mineralisation would reveal itself experimentally by a peak around $c_{max}$. Since this is not observed in measurements of trabecular BMDDs, including in situations of low turnover, we conclude that mineralisation ends smoothly by an ever-reducing mineralisation rate, so that secondary mineralisation continues over a time scale that is commensurate with the human lifespan.

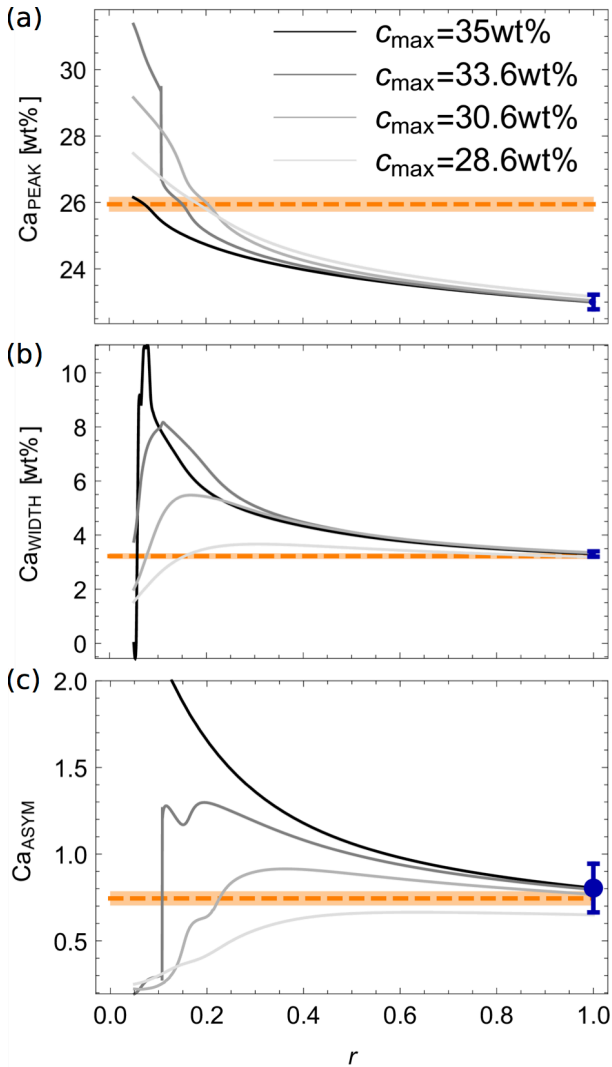

**Figure 4** – Shape parameters of the simulated BMDDs as a function of turnover rate $r$ (solid grey lines) and maximum calcium capacity $c_{max}$. The shape parameters measured on the reference trabecular BMDD are shown in blue with error bars corresponding to standard deviation. The values measured on the average femoral cortical BMDD are shown as orange dotted lines, with standard deviations shown as orange shaded areas.

3) Our simulations shows that a decrease in bone turnover generally results in a shift of $Ca_{PEAK}$ towards higher calcium densities while the BMDD peak broadens. The limitation of the amount of calcium that can be incorporated into bone results in an interesting constraint on the different shape parameters of the BMDD. With decreasing bone turnover, not only the width of the peak changes from broadening to narrowing, but also the asymmetry of the peak reverses its behaviour (Fig. 4c). This suggests that parameters characterising the asymmetry of the BMDD peak should be given more attention in future BMDD studies.

Since our model calculations are based on the trabecular reference BMDD of healthy individuals, we were careful in restricting our conclusions not only to adults, but also to trabecular bone. However, the structural organisation of human trabecular bone is known to be similar to cortical bone on the micro and nano scales [16]. On the microscale, bone is organised in both bone types in lamellae. On the nanoscale, bone constitutes a nano-composite of an organic collagen matrix and inorganic bone apatite particles. The micro and nano scales are the length scales on which processes of remodelling and mineralisation cause structural changes. Consequently, we find it acceptable to address with our last conclusion cortical bone.

4) Imposing lower maximum calcium capacities did not enable us to understand the difference between the experimental trabecular reference BMDD and femoral cortical BMDDs as a difference in turnover rate only, as it is often hypothesised [8]. No value of maximum calcium capacity and turnover rate reduction was able to match jointly peak location, peak width, and peak asymmetry of the femoral cortical BMDD. Specifically, the BMDD peak was either too broad, as quantified by the peak width, $Ca_{WIDTH}$, or too asymmetric, as quantified by a measure of asymmetry of the peak, $Ca_{ASYM}$ (Fig. 4). Table 1 shows that the shape of the BMDD in healthy adults is the same in trabecular bone and femoral cortical bone, but the peak is located at lower values of calcium content in trabecular bone. Confronting this with the pluralism of simulated BMDD shapes in Figure 2, the similarity of the experimental BMDD shapes between trabecular and cortical bone is remarkable. How these two different bone types can exhibit such a similar structural heterogeneity remains an open question, but it suggests that additional mechanisms play a role in regulating bone's tissue material properties. In particular, bone resorption has been assumed to target any calcium content indistinctly in the present work, but resorption patterns may differ between trabecular bone, where bone resorption predominantly removes lowly mineralised bone near trabecular surfaces, and cortical bone, where bone of any calcium density may be reached by tunnelling basic multicellular units. This will be the subject of future works.

## Declaration of interest
The authors report no conflicts of interest. The authors alone are responsible for the content and writing of the paper.

## Acknowledgment
CL is funded by the Monash University Postgraduate Publications Award. PRB and RW acknowledge support by Universities Australia (UA) and German Academic Exchange Service (DAAD) for the Australia–Germany Joint Research Co-operation Award (Project "The cellular control of mineral heterogeneity in bone").